\documentclass{article}
\newcommand{\BE}{\begin{equation}}
\newcommand{\EE}{\end{equation}}
\newcommand{\BA}{\begin{eqnarray}}
\newcommand{\EA}{\end{eqnarray}}

\input epsf.tex

\begin{document}

%\twocolumn[\hsize\textwidth\columnwidth\hsize\csname@twocolumnfalse\endcsname
\title{{\normalsize \bf Acoustic scattering by a cylinder near a pressure release
surface}}
\author{{\small Zhen Ye and You-Yu Chen}\\
{{\footnotesize \it Department of Physics, National Central
University, Chungli, Taiwan 32054, Republic of China}}}

\date{\small (\today)  }
\maketitle %\draft

\begin{abstract}

\noindent This paper presents a study of acoustic scattering by a
cylinder of either infinite or finite length near a flat
pressure-release surface. A novel self-consistent method is
developed to describe the multiple scattering interactions between
the cylinder and the surface. The complete scattering amplitude
for the cylinder is derived from a set of equations, and is
numerically evaluated. The results show that the presence of the
surface can either enhance or reduce the scattering of the
cylinder, depending on the frequency, the composition of the
cylinder, and the distance between the cylinder and the surface.
Both air-filled and rigid cylinders are considered.
\\
\vspace{1pt} 
\vspace{1pt}
\end{abstract}
%\pacs{PACS numbers: 43.20.Fn, 43.30.Gv., 43.30.Bp.}
%]

\section*{\small \bf INTRODUCTION}

Acoustic scattering by underwater objects near a pressure release
boundary is a very important issue in a number of current research
and applications, including the modeling of scattering from
surface dwelling fish, the understanding of oceanic fluxes and
ambient noises generated at ocean surface layers. It may also be
of great help in models of acoustic scattering by submarines near
the ocean surface.

In the literature, the research on sound scattering by underwater
objects near a pressure release surface has been mainly focused on
the scattering by a spherical object such as an air bubble (Refs.
e.~g. \cite{Strasburg,Oguz,Tolstoy,Huang,Strasburg2,Huang1,Ye1}).
In many important applications, however, underwater objects may
not take the spherical geometry. Rather they often take elongated
shapes. This includes, for example, the surface dwelling fish, the
floating logs in rivers, military objects, and so on. For these
situations, it is desirable to study acoustic scattering by an
elongated object near a boundary. By searching the literature, we
find that the research along this line is surprisingly scarce. The
purpose of the the present paper is to present an investigation of
acoustic scattering by a cylinder of either infinite or finite
length near a flat pressure-release boundary.

We consider acoustic scattering by an elongated object near a flat
pressure release surface; the sea or river surface can be regarded
as one of such surfaces when the acoustic wavelength is long
compared to the surface wave. As a first step, for simplicity yet
not to compromising the generality, we assume the object as a
straight cylinder. Due to the presence of the surface, the wave
will be scattered back and forth between the surface and the
object before it reaches a receiver. The rescattering from the
scatterer and rereflection from the surface are studied using a
self-consistent method by expressing all the waves in terms of
modal series. The scattering by the cylinder is thus exactly
evaluated, and analyzed. The theory is first developed for an
infinite cylinder, then extended to finite cylinders using the
genuine approach given by Ref.~\cite{Ye2}. Although the theory
allows us to consider a variety of cylinders, in order to show the
essence of the theory in its most transparent way we focus on two
important types of cylinders, that is, the air-filled and the
rigid cylinders. The former can be used to model the fish while
the latter may resemble some acoustic scattering characteristics
of military objects.

\section{\small \bf FORMULATION OF THE PROBLEM}

\begin{figure}
\begin{center}
\epsfxsize=4in\epsffile{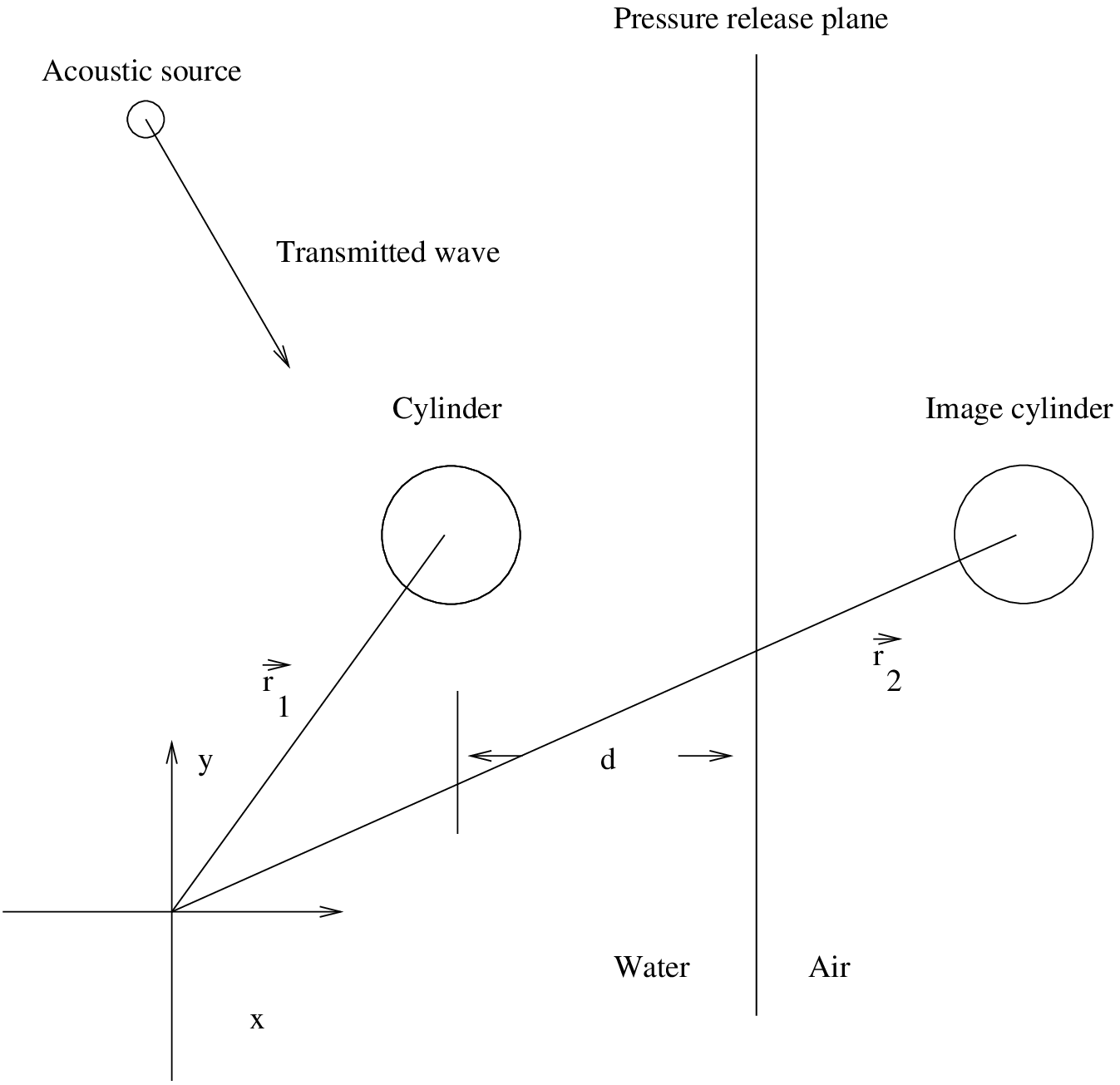} \vspace{14pt}\caption{Schematic
diagram for an cylinder near a flat pressure release surface }
\label{fig1}
\end{center}
\end{figure}

The problem considered in this paper is depicted in
Fig.~\ref{fig1}. A straight cylinder is located in the water at a
depth $d$ beneath a pressure release plane which can be the sea
surface. For simplicity, we assume that the axis of the cylinder
is parallel to the plane. The radius of the cylinder is $a$. The
acoustic parameters of the cylinder are taken as: the mass density
$\rho_1$ and sound speed $c_1$, while those of the surround water
are $\rho$ and $c$; therefore the acoustic contrasts are
$g=\rho_1/\rho$ and $h=c_1/c$. A parallel line acoustic source
transmitting a wave of frequency $\omega$ is at $\vec{r}_s$ some
distance away from the surface. The transmitted wave is scattered
by the cylinder and reflected from the surface, as shown in
Fig.~\ref{fig1}. The reflected wave is also scattered by the
cylinder. The wave scattered by the cylinder is again reflected by
the surface. Such a process is repeated, establishing an infinite
series of rescattering and rereflection between the cylinder and
the surface. This multiple scattering process can be conveniently
treated by a self-consistent manner. The rectangular frame is set
up in such a way that the $z$-axis is parallel to the axis of the
cylinder. the $x$-axis and $y$-axis are shown in Fig.~\ref{fig1}.
To solve the scattering problem, however, we use the cylindrical
coordinates in the rectangular system. We note that in the present
paper, for brevity we do not consider the case that the incident
direction is oblique to the axis of the cylinder; the extension to
oblique cases is straightforward. The setting in the problem is by
analogy with that described in Ref.~\cite{Ye1}, where a spherical
air bubble is placed beneath the flat boundary.

\subsection{Scattering by a cylinder of infinite length}

In this section, we present a formulation for sound scattering by
an infinite cylinder near a pressure-release boundary. For
succinctness, we only show the most essential steps in the
derivation. First the direct wave from the line source can be
written as \BE p_{inc}=i\pi
H_0^{(1)}(k|\vec{r}-\vec{r}_s|),\label{eq:pinc}\EE with $k$ being
the wave number of the transmitted wave ($k=\omega/c$), and
$H_0^{(1)}$ being the zero-th order Hankel function of the first
kind. The reason why we choose to use the line source is that it
can easily used to include the usual plane wave situation; for
this we just need to put the source at a place so that
$k|\vec{r}-\vec{r}_s|
>>1$. Due to the presence of the pressure release surface, the reflection from the
surface of the direct wave can be regarded as coming from an image
source located symmetrically about the surface, and is written as
\BE p_r = -i\pi H_0^{(1)}(k|\vec{r}-\vec{r}_{si}|),
\label{eq:pr}\EE where $\vec{r}_{si}$ is the vector coordinate for
the image, which is at the parity position about the plane.

The scattered wave from the cylinder can be generally written as
\BE p_{s1} = \sum_{n=-\infty}^{\infty}
A_nH_n^{(1)}(k|\vec{r}-\vec{r}_1|)e^{in\phi_{\vec{r}-\vec{r}_1}},
\label{eq:ps1} \EE where $A_n$ are the coefficients to be
determined later, $H_n^{(1)}$ are the $n$-th order Hankel
functions of the first kind, and $\phi$ is the azimuthal angle
that sweeps through the plane perpendicular to the longitudinal
axis of the cylinder. According to Brekhovskikh\cite{Bres}, the
effect of the boundary on the cylinder can be represented by
introducing an image cylinder located at the mirror symmetry site
about the plane surface. The rereflection and rescattering between
the surface and the cylinder can be represented by the multiple
scattering between the cylinder and its image. The scattered wave
from this image can be similarly written as \BE
p_{s2}=\sum_{n=-\infty}^{\infty}
B_nH_n^{(1)}(k|\vec{r}-\vec{r}_2|)e^{in\phi_{\vec{r}-\vec{r}_2}},
\label{eq:ps2}\EE where $\vec{r}_2$ is the location of the image
of the cylinder, which is symmetric about the pressure-release
plane. At the pressure release surface, the boundary condition
requires $p_{s1}+p_{s2}=0$, leading to \BE B_n =
-A_{-n},\label{eq:BA} \EE where we have used the relations $$
\phi_{\vec{r}-\vec{r}_1} = \pi - \phi_{\vec{r}-\vec{r}_2}, \ \
\mbox{and}\  \ H_{n}^{(1)}(x) = (-1)^n H_{-n}^{(1)}(x). $$
Similarly the wave inside the cylinder can be written as \BE
p_{in}= \sum_{n=-\infty}^{\infty} C_n
J_n(k|\vec{r}-\vec{r}_1|)e^{in\phi_{\vec{r}-\vec{r}_1}}.
\label{eq:pins}\EE Again, $C_n$ are the unknown coefficients, and
$J_n$ are the $n$-th order Bessel functions of the first kind.

To solve for the unknown coefficients $A_n$ (thus $B_n$) and
$C_n$, we employ the boundary conditions at the surface of the
cylinder. For the purpose, we express all wave fields in the
coordinates with respect to the position of the cylinder. This can
be achieved by using the addition theorem for the Hankel functions
\BE H_n^{(1)}(k|\vec{r}-\vec{r}'|)e^{in\phi_{\vec{r}-\vec{r}'}} =
e^{in\phi_{\vec{r}_1-\vec{r}'}}\sum_{l=-\infty}^{\infty}H_{n-l}^{(1)}(k|\vec{r}_1-\vec{r}'|)
e^{-il\phi_{\vec{r}_1-\vec{r}'}}
J_l(k|\vec{r}-\vec{r}_1|)e^{il\phi_{\vec{r}-\vec{r}_1}},
\label{eq:addition} \EE where $\vec{r}'$ can either be the
location of the source by setting $\vec{r}' = \vec{r}_s$, the
location of the image of the source with $\vec{r}'= \vec{r}_{si}$,
or the location of the image of the cylinder with
$\vec{r}'=\vec{r}_2$. The boundary conditions on the surface of
the cylinder state that both the acoustic field and the radial
displacement be continuous across the interface. Applying the
addition theorem to the expressions for the concerned waves in
Eqs.~(\ref{eq:pinc}), (\ref{eq:pr}), (\ref{eq:ps2}), and
(\ref{eq:pins}), then plugging them into the boundary conditions,
and after a careful calculation, we are led to the following
equation \BE D_l - \sum_{n=-\infty}^\infty A_{-n}
e^{i(n-l)\phi_{\vec{r}_1-\vec{r}_2}}H_{n-l}^{(1)}(k|\vec{r}_1-\vec{r}_2|)
=  \Gamma_l A_l, \label{eq:solution}\EE where we have used $$ B_n
= -A_{-n}.$$ In Eq.~(\ref{eq:solution}), we derived \BE \Gamma_l
=- \frac{{ H_l^{(1)}(ka) J_l'(ka/h) - gh H_l^{(1)}}'(ka)J_l(ka/h)
}{J_l(ka)J_l'(ka/h)-ghJ_l'(ka)J_l(ka/h)},\label{eq:Gamma} \EE and
\BE  D_l =
i\pi\left[H_{-l}^{(1)}(k|\vec{r}_1-\vec{r}_s|)e^{-il\phi_{\vec{r}_1-\vec{r}_s}}
-
H_{-l}^{(1)}(k|\vec{r}_1-\vec{r}_{si}|)e^{-il\phi_{\vec{r}_1-\vec{r}_{si}}}\right].
\EE

The coefficients $A_n$ are thus determined by a set of
self-consistent equations in (\ref{eq:solution}). Once $A_n$ are
found, the total scattered wave can be evaluated from \BA p_s &=&
p_{s1} + p_{s2} \nonumber\\ &=& \sum_{n=-\infty}^{\infty}
\left[A_n H_n^{(1)}(k|\vec{r}-\vec{r}_1|)e^{i n
\phi_{\vec{r}-\vec{r}_1}} + B_nH_n^{(1)}(k|\vec{r}-\vec{r}_2|)e^{i
n\phi_{\vec{r}-\vec{r}_2}}\right]. \EA In the far field limit, $r
\rightarrow \infty$, by expanding the Hankel functions, we have
\BA p_s &\approx& \sqrt{\frac{2}{\pi r}} e^{ikr} \sum_{n =
-\infty}^\infty e^{-i(n\pi/2+\pi/4)} \left[A_n
e^{-ik\vec{r}_1\cdot\hat{r}} + B_n
e^{-ik\vec{r}_2\cdot\hat{r}}\right]e^{in\phi_{\vec{r}}}\nonumber
\\
&=& \sqrt{\frac{2}{\pi r}} Q e^{ikr},\EA where we define \BE Q
\equiv \sum_{n = -\infty}^\infty e^{-i(n\pi/2+\pi/4)} \left[A_n
e^{-ik\vec{r}_1\cdot\hat{r}} + B_n
e^{-ik\vec{r}_2\cdot\hat{r}}\right]e^{in\phi_{\vec{r}}},
\label{eq:Q}\EE with $B_n = -A_{-n}$, as a measure of the
scattering strength.

\subsection{Scattering by a cylinder of finite length}

In practice, we are often concerned with acoustic scattering by
objects of finite length. Here we consider the scattering by a
finite cylinder beneath a flat pressure release surface such as
the sea plane. The problem of acoustic scattering by a finite
object has been difficult enough, let alone the presence of a
boundary. Exact solutions only exist for simply shaped objects.
Approximate methods have been developed. A review on various
methods for computing sound scattering by an isolated elongated
object is presented in Ref.~\cite{Ye2}. In this section, we extend
the cylinder-method proposed in Ref.~\cite{Ye2}, devised for an
isolated cylinder, to the present case of a cylinder near a
boundary. The reason for choosing this method is that it has been
verified both theoretically and experimentally that this method is
reasonably accurate for a wide range of situations\cite{Ye3,Ding}.
This is particularly true for the scenarios discussed in the
present paper.

From the Kirchhoff integral theorem, the scattering function from
any scatter can be evaluated from \BE f(\vec{r}, \vec{r}_i) =
-\frac{e^{-ik\vec{r}_1\cdot\hat{r}}}{4\pi}\int_{S} ds'
e^{-ik\vec{r}'\cdot\hat{r}}\vec{n}\cdot\left[\nabla_{r'}p_s(\vec{r}')
+ ik\hat{r}p_s(\vec{r}')\right], \label{eq:int}\EE where $\vec{n}$
is an outwardly directed unit vector normal to the surface, and
$\hat{r}$ is the unit vector in the scattering direction defined
as $\hat{r}=\vec{r}/r$. Function $f(\vec{r}, \vec{r}_i)$ refers to
the scattering function for incident direction at $\vec{r}_i$
implicit in the scattering field $p_s(\vec{r})$ and the scattering
direction $\hat{r}$.

First we consider the scattering from the cylinder. Then in
Eq.~(\ref{eq:int}), the field $p_s$ is the scattering field taking
values at the surface of scatterer. According to \cite{Ye2}, this
can be mimicked by that of an infinite cylinder of the same
radius. On the surface of the cylinder (not the image), from
Eq.~(\ref{eq:ps1}) the scattered field can be expressed as \BE
p_{s1}= \sum_{n=-\infty}^\infty A_nH_n^{(1)}(ka)e^{in\phi}, \EE
and \BE \vec{n}\cdot\nabla_{r'}p_{s1}= \sum_{n=-\infty}^\infty
A_nk{H_n^{(1)}}'(ka)e^{in\phi}. \EE Then the integral for the
scattering function of the cylinder, using Eq.~(\ref{eq:int}),
becomes \BE f^c(\vec{r}, \vec{r}_i) = \sum_{n=-\infty}^\infty
f_n(\vec{r}, \vec{r}_i), \EE with \BA f_n(\vec{r}, \vec{r}_i)&=&
\frac{-aL A_n e^{-ik\vec{r}_1\cdot\hat{r}}}{4\pi}\int_0^{2\pi}
d\phi e^{-ika\cos(\phi_{scat}-\phi)}\nonumber\\ & &\times
\left[ik\cos(\phi_{scat}-\phi) H_n^{(1)}(ka)e^{in\phi} +
k{H_n^{(1)}}'(ka)e^{in\phi}\right], \label{eq:fn}\EA where
$\phi_{scat}$ is the scattering angle with respect to $x-$axis
(i.~e. $\phi_{scat}=\phi_{\vec{r}}$).

Using integral identities \BE \int_0^{2\pi} d\phi e^{-ika\cos(\phi
- \phi_{scat})} e^{in\phi} = 2\pi(-i)^nJ_n(ka)e^{in\phi_{scat}},
\EE and \BE \int_0^{2\pi} d\phi e^{-ika\cos(\phi -
\phi_{scat})}\cos(\phi - \phi_{scat}) e^{in\phi} =
2\pi(-i)^niJ_n'(ka)e^{in\phi_{scat}}, \EE we can reduce
Eq.~(\ref{eq:fn}) to \BE f_n(\vec{r}, \vec{r}_i) =
\frac{-kaL(-i)^nA_ne^{-ik\vec{r}_1\cdot\hat{r}}}{2}
e^{in\phi_{scat}}\left[{H_n^{(1)}}(ka)' J_n(ka) - H_n^{(1)}(ka)
J_n'(ka)\right]. \label{eq:int2}\EE By the Wronskian identity
\BE
[J_n(x){H_n^{(1)}}'(x) - J_n'(x)H_n^{(1)}(x)] = \frac{2i}{\pi x},
\EE Eq.~(\ref{eq:int2}) becomes \BE f_n(\vec{r}, \vec{r}_i) =
\frac{-i(-i)^n L A_n e^{-ik\vec{r}_1\cdot\hat{r}}}{\pi}
e^{in\phi_{scat}}.\EE The scattering from the image of the
cylinder can be considered in the same spirit. We thus obtain \BE
f^i(\vec{r}, \vec{r}_i) = \sum_{n=-\infty}^\infty\frac{-i(-i)^nL
B_n e^{-ik\vec{r}_2\cdot\hat{r}}}{\pi} e^{in\phi_{scat}}.\EE

The total scattering function is \BA f(\vec{r}, \vec{r}_i) &=&
\sum_{n=-\infty}^\infty \left[ \frac{(-i)^{n+1}L A_n
e^{-ik\vec{r}_1\cdot\hat{r}}}{\pi} + \frac{(-i)^{n+1}L B_n
e^{-ik\vec{r}_2\cdot\hat{r}}}{\pi}\right] e^{in\phi_{scat}}
\nonumber\\ &=& \sum_{n=-\infty}^\infty \left(A_n
e^{-ik\vec{r}_1\cdot\hat{r}} +  B_n
e^{-ik\vec{r}_2\cdot\hat{r}}\right)\frac{(-i)^{n+1}L
e^{in\phi_{scat}}}{\pi}.\EA The {\it reduced} differential
scattering cross section is \BE \sigma(\vec{r}, \vec{r}_i) =
|f(\vec{r}, \vec{r}_i)/L|^2.\EE The reduced target strength is
evaluated from \BE \mbox{TS} = 10\log_{10}(\sigma).
\label{eq:TS}\EE This equation bears much similarity with the
scattering strength for the infinite cylinder given in
Eq.~(\ref{eq:Q}). In the following section, we should compute the
target strength for finite cylinders near a pressure release
boundary. In particularly, we are interested in the situation of
backscattering, in which the scattering direction is opposite to
the incident direction, i.~e. $\vec{r}=-\vec{r}_i$.

\section{\small \bf NUMERICAL RESULTS}

Some interesting properties are found for acoustic scattering by a
cylindrical object beneath a flat pressure release plane. Two
kinds of cylinders are considered: air-filled and rigid cylinders.

\begin{figure}
\begin{center}
\epsfxsize=4in\epsffile{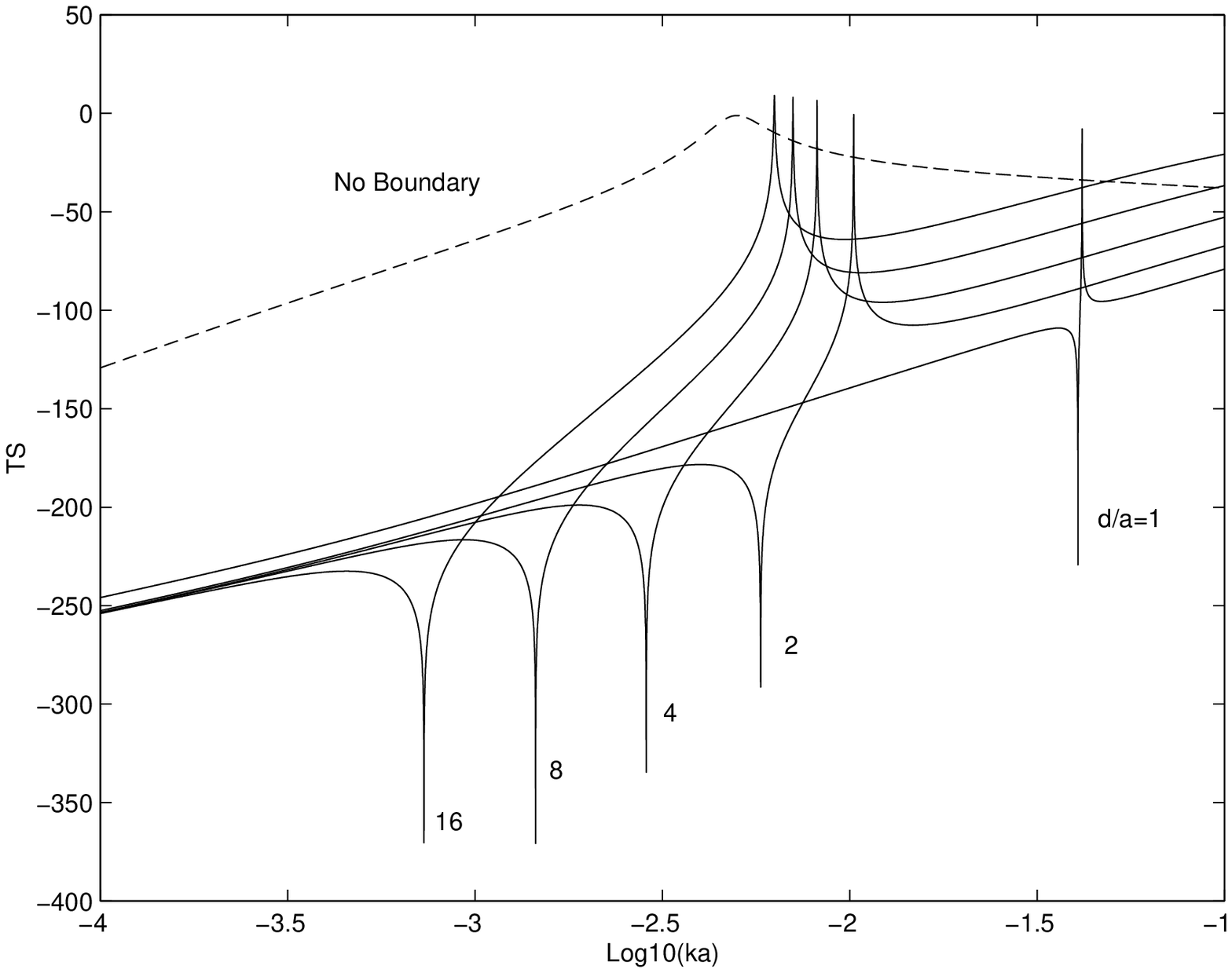} \vspace{14pt}\caption{Air
Cylinder: Backscattering target strength versus frequency for
various $d/a$ values. The incident angle is $\pi/4$.} \label{fig2}
\end{center}
\end{figure}

Let us first consider the sound scattering by an air-filled
cylinder of length $L$. Although the theory developed in the last
section allows the study of scattering for arbitrary incident and
scattering angles, we will first concentrate on backscattering. In
addition, without notification we will consider the incident at an
angle of $\pi/4$ with respect to the normal to the flat surface.
Fig.~\ref{fig2} shows the reduced backscattering target strength
in an arbitrary unit as a function of frequency in terms of the
non-dimensional parameter $ka$. The cylinder is placed at the
depths of $d/a = 1, 2, 4, 8$, and $16$ respectively. For
comparison, the situation that the boundary is absent is also
plotted. Without boundary, the scattering by a single cylinder has
a resonant peak at about $ka=0.005$. When a flat pressure-plane is
added, the scattering from the cylinder will be greatly suppressed
for most frequencies under consideration, except for the
resonance. At the resonance, the scattering is in fact enhanced by
the presence of the surface. This is a unique feature for the
cylinder situation. Another effect of the boundary is to shift the
resonance peak of the cylinder towards higher frequencies. As the
distance between the cylinder and the surface is decreased, the
position of the peak moves further towards higher frequencies, and
the resonance peak is becoming narrower and narrower. Before the
resonance peak, there is a prominent dip in the scattering
strength. For the extreme case that the cylinder touches the
boundary, the significant dip appears immediately before the
resonance. This dip is not observed in the case of a spherical
bubble beneath a boundary\cite{Ye1}.

\begin{figure}
\begin{center}
\epsfxsize=4in\epsffile{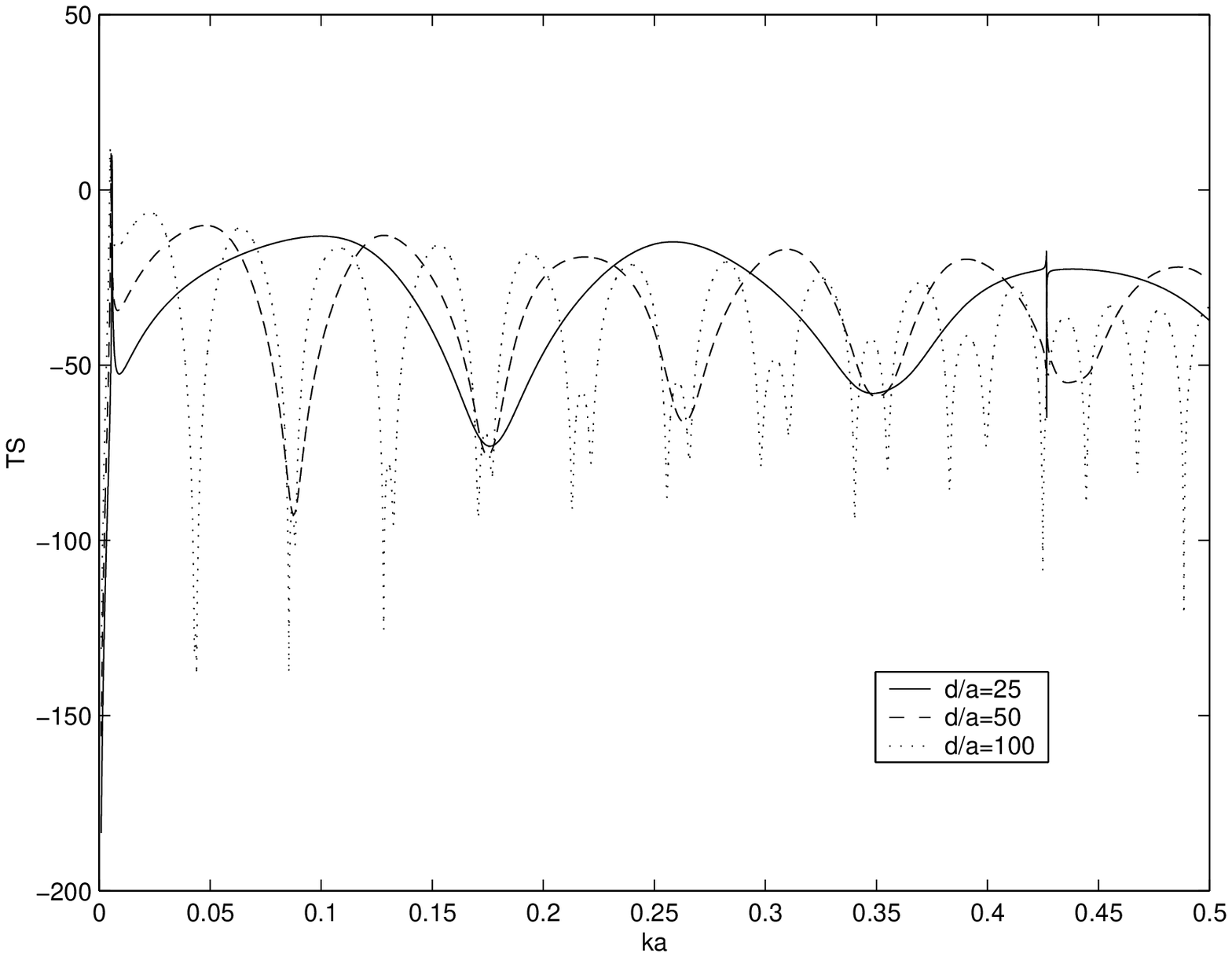} \vspace{14pt}\caption{Air
Cylinder: Backscattering target strength versus frequency for
larger $d/a$ values. The incident angle is $\pi/4$.} \label{fig3}
\end{center}
\end{figure}

When the distance between the cylinder and the surface is
increased, the resonance peak moves to lower frequencies until
reaching that of the cylinder without a boundary. In
Fig.~\ref{fig3}, the reduced target strength is plotted against
$ka$ for $d/a = 25, 50,$ and 100. Here we see that, as the
cylinder is moved further from the surface, regular oscillatory
features appear in the scattering strength around the values
without the boundary. The observed peaks and nulls are mainly due
to interference effects between the cylinder and the boundary, as
these oscillatory features persist even when the multiple
scattering is turned off. The nulls, appearing at some frequency
intervals, are more numerous and are spaced more closely together
as the cylinder is moved away from the boundary. The peak and null
structures are somewhat in accordance with the Lloyd's mirror
effect. These features are in analogy with the results shown for
the case of a spherical bubble beneath the boundary \cite{Ye1}.
However, there is a distinct difference. Namely, the separation
between the peaks or between the nulls decreases as the frequency
increases.

\begin{figure}
\begin{center}
\epsfxsize=4in\epsffile{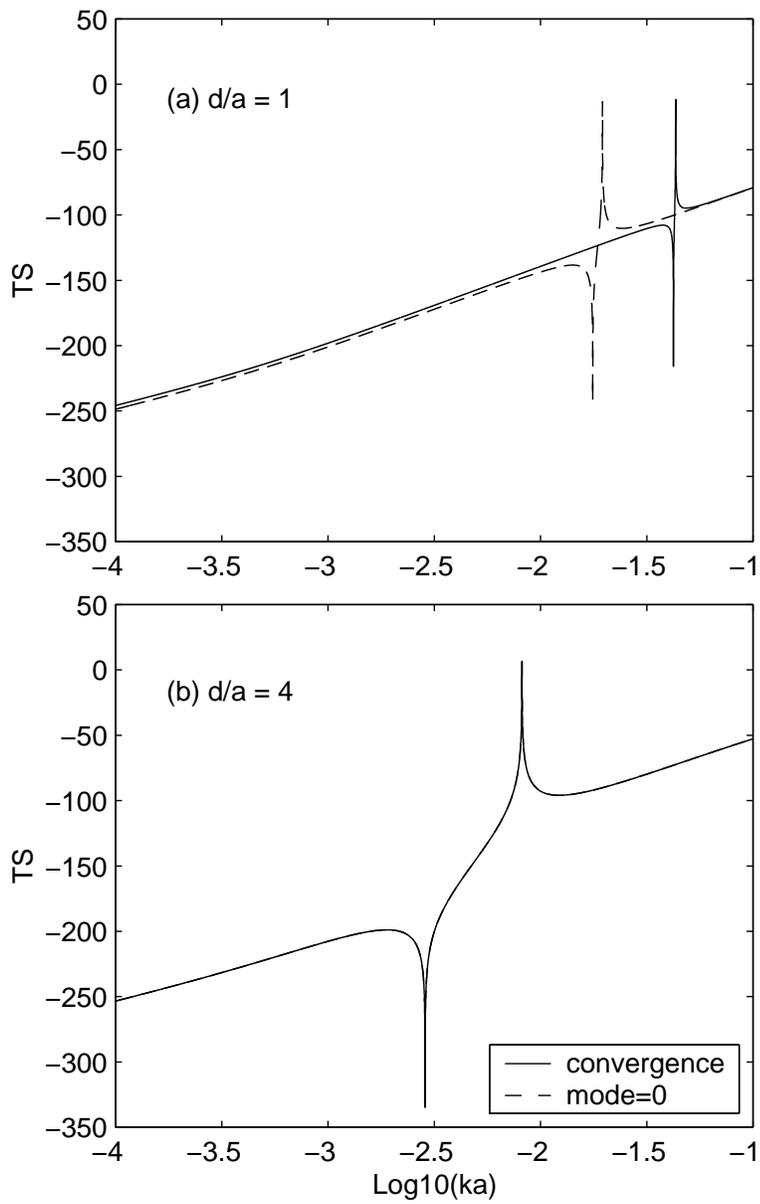} \vspace{14pt}\caption{Air
Cylinder: Backscattering target strength versus frequency for
different modes. The incident angle is $\pi/4$.} \label{fig4}
\end{center}
\end{figure}

We have also studied the contributions from different oscillation
modes of a cylinder to the scattering. From Eq.~(\ref{eq:TS}), it
is clear that the scattering is contributed from various vibration
modes and the contributions are represented by the summation in
which the index $n$ denotes the modes. We find that when the
cylinder is located far enough from the surface, the scattering is
dominated by $n=0$ mode for low frequencies (e.~g. $ka<1$); mode
$n=0$ is the omni-directional pulsating mode of the cylinder,
i.~e. its scattering is uniform in every direction. When the
cylinder is moved close to the surface, higher vibration modes
become important. These properties are illustrated in
Fig.~\ref{fig4}. For the extreme case that the cylinder touches
the boundary as shown in Fig.~\ref{fig4}(a), the result from
including only $n=0$ mode is compared with that including all
modes. It is interesting to see that the effect of coupling the
pulsating mode with other modes is only to shift the resonance and
dip peaks. For low frequencies away from the resonance and the
dip, the effect from higher models is not evident. As the cylinder
is move away from the surface, the effect of higher modes
gradually decreases. For the case $d/a=4$, the effect of higher
modes (i.~e. $|n|\geq 2$) virtually diminished.

\begin{figure}
\begin{center}
\epsfxsize=4in\epsffile{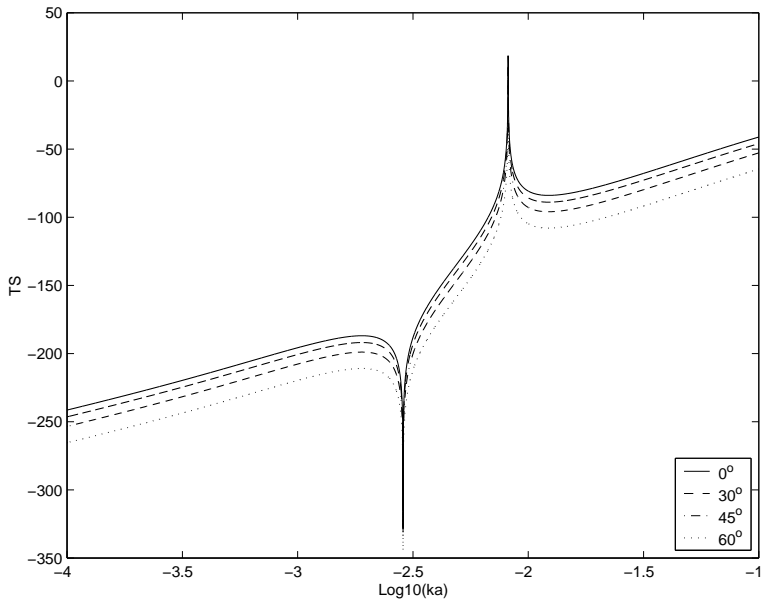} \vspace{14pt}\caption{Air
Cylinder: Backscattering target strength versus frequency for
various incident angles. The incidence angle is measured with
respect to the $x$-axis, referring to Fig.~\ref{fig1}. Here $d/a =
4$.} \label{fig5}
\end{center}
\end{figure}

The effects of the incident angle on the back scattering is shown
by Fig.~\ref{fig5}. The results show that the scattering is highly
anisotropic except at the scattering dip and peak positions; note
the scale used in plotting Fig.~\ref{fig5}. The fact that the
scattering dip does not rely on the incident angle implies that it
is not caused by the Lloyd mirror effect. This is because if it
were due to the Lloyd mirror effect, different incident angles
would lead to different acoustic paths in reflection and incidence
and thus result in different phases, causing the scattering
pattern to vary.

\begin{figure}
\begin{center}
\epsfxsize=4in\epsffile{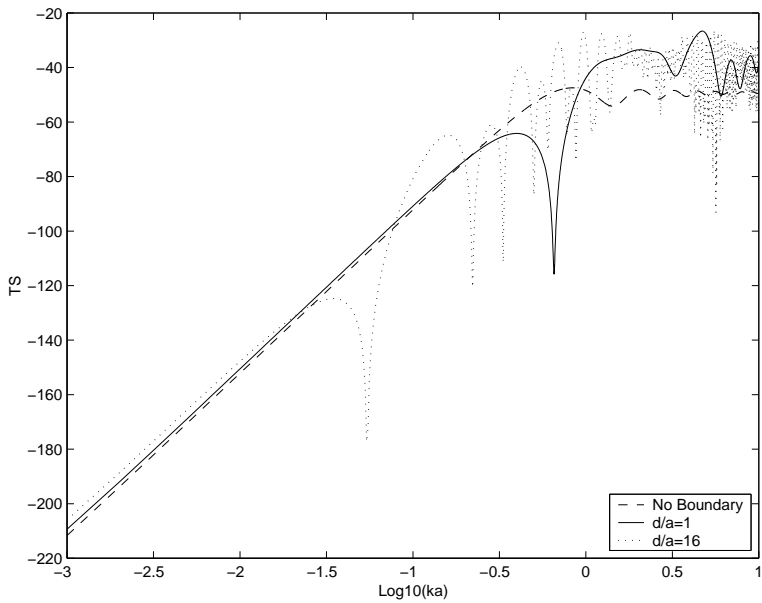} \vspace{14pt}\caption{Rigid
cylinder: Backscattering target strength versus frequency for
various $d/a$. The incident angle is $\pi/4$.} \label{fig6}
\end{center}
\end{figure}

Next we consider scattering from a rigid cylinder beneath a
pressure release boundary. For the rigid cylinder, in contrast to
the air cylinder case, the scattering is not so significantly
reduced by the presence of the surface. Instead, it is interesting
that the presence of the surface in fact can enhance the
scattering strength for most frequencies, except for the
frequencies at which the Lloyd effect comes into function. This
enhancement is particularly obvious in the low frequency regime.
Similar to the air cylinder case, when the distance is large
enough, the Lloyd mirror effect causes the scattering strength to
oscillate around the values without the boundary for low
frequencies. Fig.~\ref{fig6} shows that for low frequencies, the
frequency dependence of the scattering is similar for different
distances between the cylinder and the surface. For high
frequencies, e.~g. $ka >0.4$, the multiple scattering is evident
and is shown to increase the scattering strength.

\begin{figure}
\begin{center}
\epsfxsize=4in\epsffile{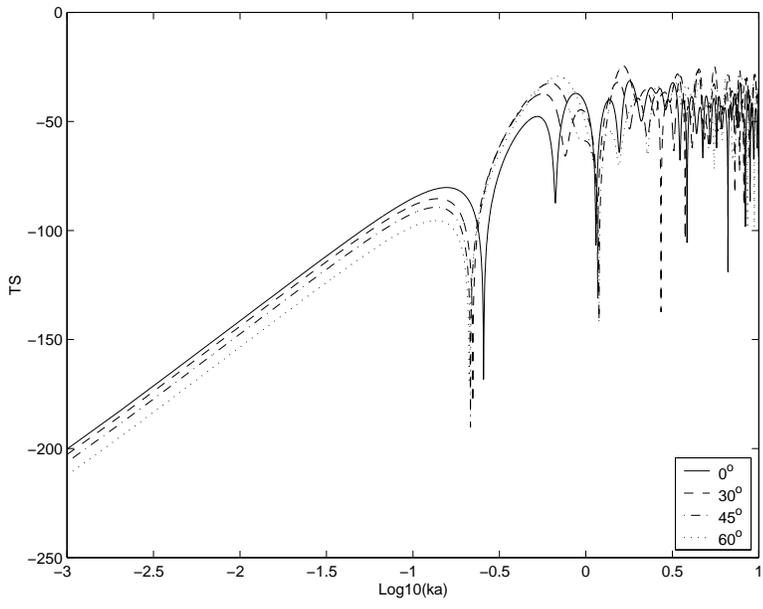} \vspace{14pt}\caption{Rigid
cylinder: Backscattering target strength versus frequency for
various incident angles with $d/a = 4$.} \label{fig7}
\end{center}
\end{figure}

The backscattering by the rigid cylinder under the boundary is
anisotropic. This is illustrated in Fig.~\ref{fig7}, which shows
the backscattering target strength as a function of $ka$ for
different incidence angles. The separation between he cylinder and
the surface is $d/a =4$, and the incidence angle is measured with
respect to the $x$-axis, referring to Fig.~\ref{fig1}. For low
frequencies, i.~e. $ka < 0.1$, the scattering is strongest when
the incidence is normal to the surface (i.~e. for the zero degree
incidence). Different from the above air cylinder case, the dips
in the scattering strength depend on the incident angles.

\begin{figure}
\begin{center}
\epsfxsize=4in\epsffile{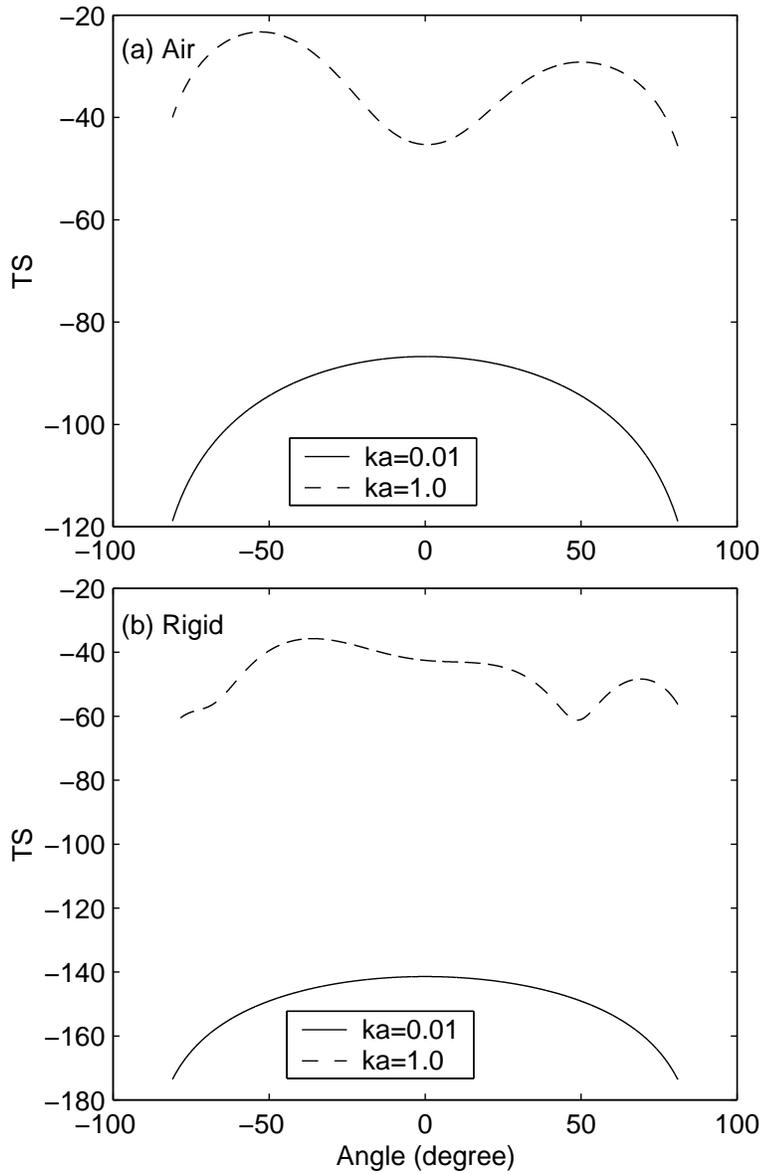} \vspace{14pt}\caption{Bistatic
scattering target strength versus scattering angle for two
frequencies $ka = 0.01, 0.1$: (a) Air-filled cylinder, (b) Rigid
cylinder. Here $d/a = 4$ and the incidence angle is 45 degree. The
scattering angle is measured with respect to the negative $x$-axis
referring to Fig.~\ref{fig1} } \label{fig8}
\end{center}
\end{figure}

Finally we consider the bistatic scattering. The scattering is in
the $x-y$ plane (See Fig.~\ref{fig1}). We fix the incident angle
at $45$ degree with respect to the normal to the boundary. The
scattering azimuthal angle is measured from the negative direction
of the $x$-axis (Referring to Fig.~\ref{fig1}). Fig.~\ref{fig8}
shows the scattering angle dependence of the bistatic scattering
target strength for the air filled and rigid cylinders
respectively. It is interesting to see that when the frequency is
low, the scattering tends to be symmetric around the normal to the
boundary, i.~e. the zero degree scattering angle, for both the
air-filled and rigid cylinders. The scattering is strongest at the
zero scattering angles. This result indicates that when the
frequency is low, the scattering from a cylinder near a boundary
bears similar properties of the acoustic radiation from a dipole
source, independent of the incident angle. This feature seems
against the intuition at the first sight, but can be understood as
follows. The scattering from a target can be regarded as a second
source radiating waves into the space. From, for instance,
Eq.~(\ref{eq:ps1}), we know that the radiated wave consists of the
contributions from all vibration modes of the cylinder. The mode
of $n=0$ is the monopole which radiates an omni-directional wave.
At low frequencies, this monopole radiation dominates. In the low
frequency regime, both the cylinder and its image radiate waves
but in the opposite phase. If the monopole mode dominates, the
resulting radiation should appear as that from a dipole source:
the strongest radiation is along the dipole axis. This is in fact
exactly what is shown by Fig.~\ref{fig8}. Comparing
Figs.~\ref{fig5} with \ref{fig7}, however, the fact that the
bacskscattering relies on the incident angle indicates that the
overall bistatic scattering does depend on the incident angle.
When the frequency is increased to a certain extent, the bistatic
scattering pattern is no longer symmetric around the normal to the
boundary.

\section{\small \bf SUMMARY}

In this paper, we considered acoustic scattering by cylinders near
a pressure-release boundary. A novel method has been developed to
describe the multiple scattering between the boundary and the
cylinder in terms of an infinite modal series. The complete
solution has been derived. Although the theory developed allows
for study of various cylinders, for brevity only the cases of
air-filled and rigid cylinders are considered. The numerical
results show that the presence of the boundary modifies the
scattering strength in various ways. One of the most significant
discoveries is that the present of the surface can greatly
suppress the scattering from `soft' targets while may enhance
rigid bodies. In addition, comparison has been made with the
previously investigated case of a spherical air-bubble beneath a
pressure-release boundary. The study presented here may link to
various applications such as acoustic scattering from
ocean-surface dwelling fish or from any underwater elongated
objects including submarine.

\section*{\small \bf ACKNOWLEDGEMENT}

The work received support from the National Science Council.

%\begin{references}

\newpage

\end{document}